\documentclass[11pt,a4paper]{article}
\usepackage{url}
\usepackage[margin=1in]{geometry}
\usepackage{setspace}
\usepackage{hyperref}
\usepackage{authblk}
\usepackage{setspace}
\usepackage{svg}

\usepackage{booktabs}
\usepackage{multirow}
\usepackage{tabularx}
\usepackage{array}
\usepackage{makecell}

\usepackage{graphicx}
\usepackage{caption}
\usepackage{subcaption}
\usepackage{float}

\usepackage{amsmath,amssymb,amsfonts}
\usepackage{amsthm}
\usepackage{mathrsfs}

\usepackage{algorithm}
\usepackage{algorithmicx}
\usepackage{algpseudocode}

\usepackage{listings}
\usepackage[dvipsnames]{xcolor}

\usepackage{siunitx}
\usepackage[title]{appendix}
\usepackage{textcomp}
\usepackage{mdframed}
\usepackage{tcolorbox}
\tcbuselibrary{skins,breakable}
\usepackage{empheq}
\usepackage{fancybox}
\usepackage[textsize=small]{todonotes}

\usepackage[numbers,sort&compress]{natbib}
\theoremstyle{plain}

\theoremstyle{definition}

\title{\textbf{FlowLOT: Linearized Optimal Transport for Flow Cytometry Analysis}}

\author[1]{Naqib Sad Pathan}
\author[2]{Mohammad Shifat-E-Rabbi}
\author[3]{Kristofor E. Pas}
\author[3]{Ivan Medri}
\author[4]{Bartek Rajwa}
\author[1,3,*]{Gustavo K. Rohde}

\affil[1]{Department of Electrical \& Computer Engineering, University of Virginia, Charlottesville, VA, USA}
\affil[2]{Department of Electrical and Computer Engineering, North South University, Dhaka, Bangladesh}
\affil[3]{Department of Biomedical Engineering, University of Virginia, Charlottesville, VA, USA}
\affil[4]{Bindley Bioscience Center, Purdue University, West Lafayette, IN, USA}
\affil[*]{Corresponding author: gustavo.rohde@gmail.com}

\date{} 

\begin{document}

\maketitle

\begin{abstract}

Multiparameter flow cytometry generates high-dimensional, unordered single-cell measurement data for disease diagnosis and monitoring, yet analysis often remains dependent on manual gating, limiting scalability and reproducibility. Existing machine-learning approaches can reduce annotation burden but frequently require large training cohorts and offer limited interpretability. To address these challenges, we introduce \textbf{\textit{FlowLOT}}, an optimal-transport-based framework that models the single-cell measurement data of a patient sample as an empirical cellular distribution and maps it directly into a fixed-length feature vector. Within a single transparent architecture, \textbf{\textit{FlowLOT}} unifies high-dimensional classification, interpretable visualization, and continuous quantitative inference. In few-shot regimes, using as few as 16 patients per class on FlowCAP-II and 8 patients per class on BLAST110, it accurately distinguishes healthy from acute myeloid leukemia (AML) samples, reaching $94.3 \pm 1.9\%$ and $98.0 \pm 1.7\%$ balanced accuracy, respectively. The underlying embedding exposes marker-level variation driving disease-associated population shifts and enables quantitative measurable residual disease (MRD) estimation, achieving a Pearson correlation of $0.82$ on held-out samples and $0.79$ under cross-dataset transfer. Furthermore, at the clinically relevant $0.1\%$ threshold for leukemia-associated immunophenotype (LAIP) residual disease, \textbf{\textit{FlowLOT}} detects positivity with $72\%$ sensitivity at $100\%$ specificity. By replacing subjective manual gating and black-box deep learning with a distribution-aware framework, \textbf{\textit{FlowLOT}} offers a sample-efficient, scalable, and interpretable solution for high-dimensional cytometry under realistic clinical and experimental constraints.

\end{abstract}

\textbf{Keywords:} Linearized optimal transport (LOT), High dimensional distribution (HDD), nearest subspace classifier (NSC), flow cytometry (FCM), acute myeloid leukemia (AML), measurable residual disease (MRD)

\newpage

\section*{Introduction}
Flow cytometry (FCM) is a single-cell optical analysis technology that measures light-scatter and fluorescence signals from cells in suspension, enabling rapid, multiparametric characterization of large cell populations. In cancer research, FCM has become a foundational tool for high-throughput profiling of tumor heterogeneity, immune responses, and disease-associated cellular phenotypes \cite{herzenberg2006interpreting,mckinnon2018flow, Robinson2025}. By simultaneously measuring multiple surface and intracellular markers, FCM captures cellular heterogeneity as high-dimensional, population-scale single-cell data.

Despite major advances in cytometry instrumentation, downstream analysis remains a central bottleneck. Cytometry samples are represented as an unordered set of cellular measurements, or equivalently as a high-dimensional empirical distribution \cite{gachon2025low}. Due to the unordered nature of this representation, cytometry data is not compatible with analytical frameworks that require fixed-length vector representations of samples \cite{mahan2024point}. Consequently, conventional pipelines rely on manual gating (the expert-driven process of sequentially drawing marker-based boundaries to identify cell populations) which is labor-intensive, and difficult to standardize or scale across cohorts \cite{AML_DATA}.

These challenges are especially significant in clinical applications involving acute myeloid leukemia (AML), including disease classification and measurable residual disease (MRD) monitoring. In these settings, clinically meaningful distinctions may depend on shifts in marker-expression distributions, aberrant cell populations, or altered patterns of cellular maturation rather than from large changes in dominant populations. Robust, sensitive, and reproducible analytical methods capable of capturing these distribution-level differences are therefore essential for reliable clinical decision-making \cite{BLAST110_DATA}.

Sample classification tasks, such as distinguishing healthy and pathological patients, commonly rely on a two-step pipeline \cite{hu2022application,spies2025machine}. The first step identifies cellular subpopulations within each sample, historically through manual expert gating. Computational methods, including dimensionality reduction and clustering, have been developed to automate this step. However, robust and generalizable performance remains challenging \cite{subpopulation1,subpopulation2,spies2025machine}. The second step extracts summary features from these subpopulations, such as marker-expression profiles, which are subsequently used for tasks including classification, grading, and prognosis\cite{flowbin2,swift2,aspire1,aspire2,citrus,AML_DATA,flowpeaks}. While widely adopted, this pipeline compresses samples into summary features, rather than treating them as empirical distributions. As a result, it may miss continuous phenotypic gradients, coordinated marker-expression shifts, and changes in population shape or density that are not captured by abundance-based summaries alone.


Gaussian mixture models (GMMs) offer an alternative by modeling each sample as a mixture of Gaussian components in marker space \cite{JCM,alex2024gmm}, with subpopulations represented by mixture components and classification performed using mixture-derived parameters. To accommodate the asymmetry, multimodality, and heavy tails that are characteristic of fluorescence-based cytometry distributions, several extensions of the basic Gaussian formulation have been proposed: $t$-mixtures coupled with a Box--Cox transformation \cite{lo2008tmix}, finite mixtures of multivariate skew $t$ distributions in the FLAME framework \cite{pyne2009flame}, and the Joint Clustering and Matching (JCM) procedure, in which each sample is fit by a mixture of multivariate (skew) $t$ kernels and a class-level template is constructed through random-effects modeling of inter-sample variability \cite{JCM}. This approach has been evaluated on the FlowCAP-II AML benchmark and shown to be competitive with other template- and clustering-based pipelines for AML classification.
 
A complementary line of work has sought to remove the need for \emph{a priori} specification of the number of mixture components through Bayesian non-parametric (BNP) constructions. In this direction, Dundar et al.\ developed ASPIRE, an infinite mixture of infinite Gaussian mixtures ($\mathrm{I}^{2}\mathrm{GMM}$) with Dirichlet-process priors that jointly clusters cellular events across samples while accommodating sample-specific random effects, and demonstrated near-perfect identification of anomalous AML phenotypes on the FlowCAP-II data without requiring a representative training set of AML cases \cite{aspire1}. The same framework was subsequently extended to the longitudinal monitoring of disease progression in AML, in which proportion vectors over recovered non-parametric meta-clusters were used as phenotype features for predicting the direction of change between sequential clinical samples, yielding correct calls in 90\% of relapsing cases and 100\% of stable or remitting cases on an independent test cohort \cite{Rajwa2017}.

Despite these advances, both the parametric-mixture and the BNP formulations retain important limitations when applied to complex immunophenotypes such as those encountered in AML \cite{AML_DATA}. Finite Gaussian and skew $t$ mixtures remain sensitive to initialization, model-order selection, and component identifiability, and their estimated parameters may be biased when the assumed component family is misspecified. BNP approaches mitigate the model-order problem but still impose a parametric form on each (local) component, and computational cost grows substantially with sample size and dimensionality. More fundamentally, all of these methods compress each sample into a finite-dimensional intermediate representation (either a vector of mixture parameters or a vector of meta-cluster proportions) before classification is performed. Such intermediate representations may fail to capture the full distributional structure of the underlying cytometry data, particularly for rare or transitional populations whose contribution to a global summary is small but diagnostically informative.


Existing computational approaches for MRD assessment similarly rely on intermediate clustering, heuristic decision rules, reference-based novelty detection, and varying degrees of manual intervention. In one common strategy, patient samples are jointly embedded with disease-free references, and cells absent from, or underrepresented in, the reference distribution are treated as candidate leukemic populations \cite{vial2021unsupervised,canali2023prognostic}. These candidate populations are then refined through computer-assisted manual inspection or cluster-level heuristics, such as patient-to-control cell ratios \cite{jacqmin2020clustering,weijler2022umap}. Representative methods include the GMM-based AML-MRD pipeline of Mocking et al., which identifies immature blasts using mixture models and applies reference-based novelty detection to estimate computational MRD percentages \cite{BLAST110_DATA}, as well as FlowSOM-derived approaches that use clustering to support MRD quantification \cite{bucklein2019flowsom}. These approaches can remain dependent on reference distributions, threshold choices, component or cluster definitions, and expert visual review for clinical interpretation.

Clustering- and embedding-based approaches can also be sensitive to cluster initialization, local density structure, and the rarity of leukemic cells, and may fail to resolve phenotypically overlapping populations. Dimensionality-reduction techniques such as UMAP may further distort marker relationships, potentially limiting clinical interpretability and reliability \cite{weijler2022umap}. 

More recent work has explored per-cell anomaly detection with supervised classifiers, enabling aberrancy quantification without explicit clustering and improving sensitivity to rare leukemic cells. However, these methods introduce dependence on labeled training data and visualization-guided interpretation \cite{shopsowitz2024magic}. Optimal transport-based methods offer a complementary strategy by representing samples as probability distributions and learning low-dimensional embeddings that preserve inter-patient variability relevant to MRD \cite{gachon2025low}. Yet, in existing OT-based MRD workflows, MRD inference remains indirect because it relies on quantization and post hoc clustering rather than direct estimation from the full empirical distributions. Together, these limitations motivate a unified, distribution-level framework that preserves the high-dimensional structure of cytometry data while enabling principled MRD assessment with minimal heuristic or manual intervention.

In this work, we introduce FlowLOT, a distribution-level framework for representing, classifying, and interpreting flow cytometry data. Rather than reducing each sample to manually gated populations, cluster-derived features, or mixture-model parameters, FlowLOT models each sample as a high-dimensional empirical probability distribution over single-cell marker measurements. Building on LOT theory, analytical representations of point-cloud data, and recent applications of optimal transport in single-cell analysis, we develop an LOT-based embedding tailored to cytometry measurements \cite{gachon2025low,lot_main,pointcloud3d,deeplot,bunne2023learning}.

The central motivation is to retain the full distributional structure of cytometry data while making sample-level learning computationally tractable and statistically interpretable. FlowLOT maps cytometry distributions into a linear vector space, enabling closed-form statistical learning for classification, visualization, and MRD inference—tasks that are otherwise complicated by the non-Euclidean geometry of probability distributions. This linearization supports accurate learning from limited labeled samples, a common constraint in clinical cytometry studies. Moreover, because the embedding is invertible, learned differences between groups can be mapped back to the original marker space, providing biologically interpretable descriptions of disease-associated distributional shifts. We validate FlowLOT on simulated and real-world flow cytometry datasets, demonstrating its utility for classification, visualization, and clinically relevant MRD assessment in acute myeloid leukemia.

\section*{Results}

\subsection*{Overview of the FlowLOT framework}

\begin{figure}[!htb]
    \centering \includegraphics[width=0.8\linewidth]{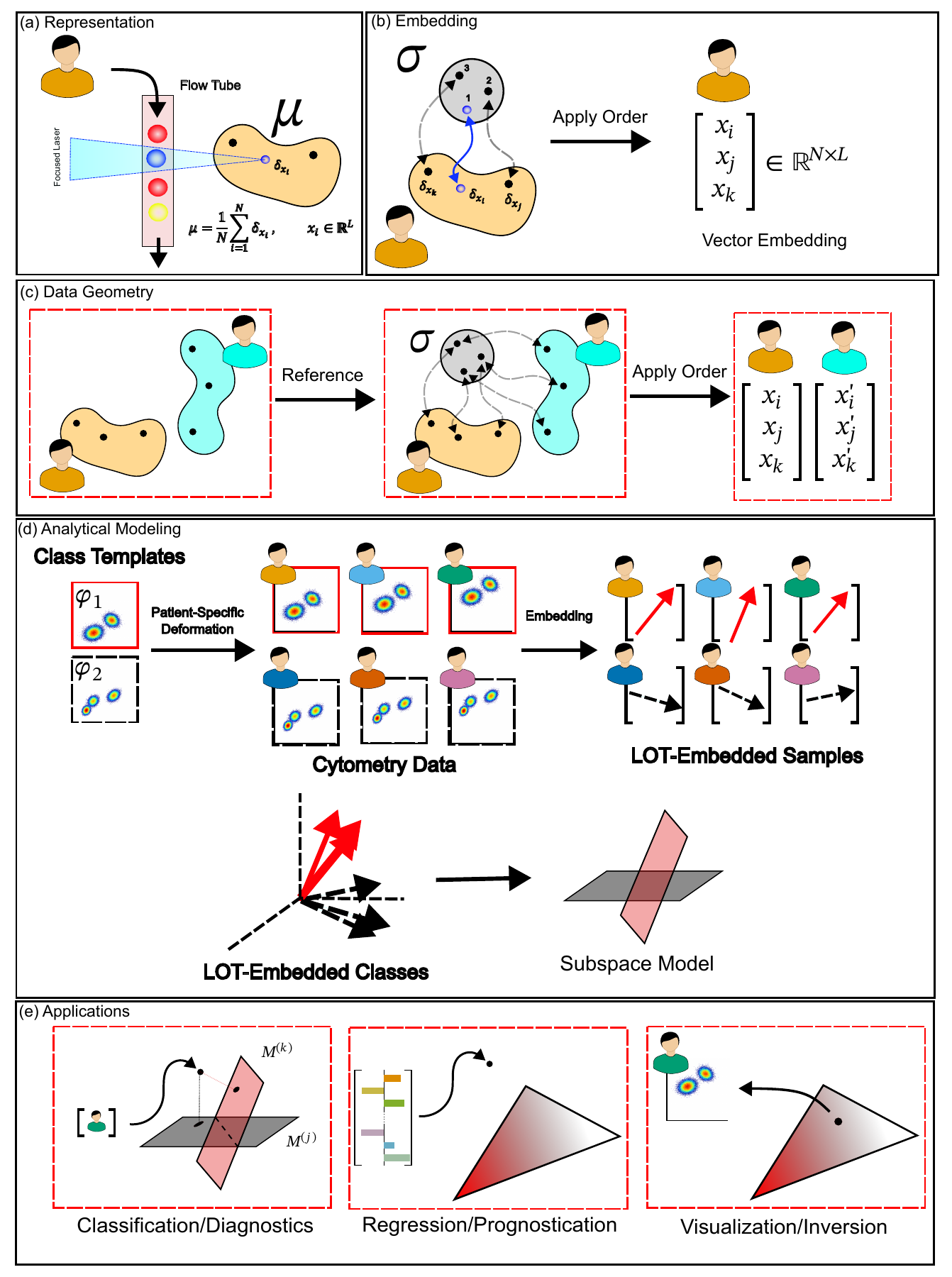}
    \caption{System diagram outlining the proposed \textbf{\textit{FlowLOT}} pipeline. (a) High-dimensional distribution (HDD)-based representation of flow cytometry data as a discrete measurement; (b) LOT-based embedding of HDD data mapping into a common-defined reference system; (c) Example showing of consistency of embedding across multiple patients into common coordinate system; (d) Analytical modeling of different patient classes generated through a deformation within their spatial domain, with embedding into a common coordinate system; (e) Downstream applications.}
    \label{fig1_overview}
\end{figure}

The proposed \textit{\textbf{FlowLOT}} framework is structured around four components as depicted in Figure ~\ref{fig1_overview} that provide a principled and interpretable approach to the analysis of high-dimensional flow cytometry data.

\paragraph{Component 1: Representation}

\textbf{\textit{FlowLOT}} represents multiparameter flow cytometry samples as discrete empirical probability distributions. Each cell is represented as a point in high-dimensional marker space, where measured parameters define the coordinate dimensions and the observed cells constitute the empirical support of the distribution. This formulation preserves within-sample cellular heterogeneity and enables principled comparison across samples using well-defined distributional distances and embeddings.

\paragraph{Component 2: Analytical class model}
Many computational approaches to flow cytometry analysis are primarily descriptive, relying on empirical summaries, clustering results, or manually defined populations rather than explicit generative models of population-level variability. The proposed \textbf{\textit{FlowLOT}} framework provides a foundation in which theoretical models of cytometry variation can be formulated and tested. Specifically, we model each phenotypic class as arising from an underlying template distribution subject to smooth, structured deformations. This assumption reflects the biological intuition that meaningful phenotypic variability manifests as coordinated shifts in cellular distributions, rather than as arbitrary, sample-specific noise.

\paragraph{Component 3: Embedding}
A central challenge in operationalizing this model is that each sample consists of an unordered, variable-size collection of single-cell measurements, making direct comparison across samples nontrivial. \textbf{\textit{FlowLOT}} addresses this challenge by introducing an optimal-transport-based embedding that maps these distribution-valued samples into fixed-length vector representations. This embedding preserves distributional geometry while enabling statistical modeling in a linear vector space.

\paragraph{Component 4: Applications}

We demonstrate the utility of the \textbf{\textit{FlowLOT}} embedding in three downstream applications. First, under the proposed generative assumptions, FlowLOT enables accurate \textbf{classification} from relatively few labeled samples. Second, it supports \textbf{visualization} of feature-wise variability between phenotypic classes, allowing disease-associated distributional shifts to be interpreted in marker space. Third, the same representation supports clinically relevant \textbf{quantification} tasks, including estimation of MRD.

\subsection*{Synthetic data demonstration}

To illustrate the geometric interpretability and discriminative power of the proposed framework, we first consider a controlled synthetic example. We generate two high-dimensional data classes, each defined by a template distribution, and construct individual samples within each class by deforming the corresponding template. This setup isolates the class-specific structure and enables direct visualization of how this structure is encoded in the embedding space.

\begin{figure}[H]
    \centering \includegraphics[width=0.7\linewidth]{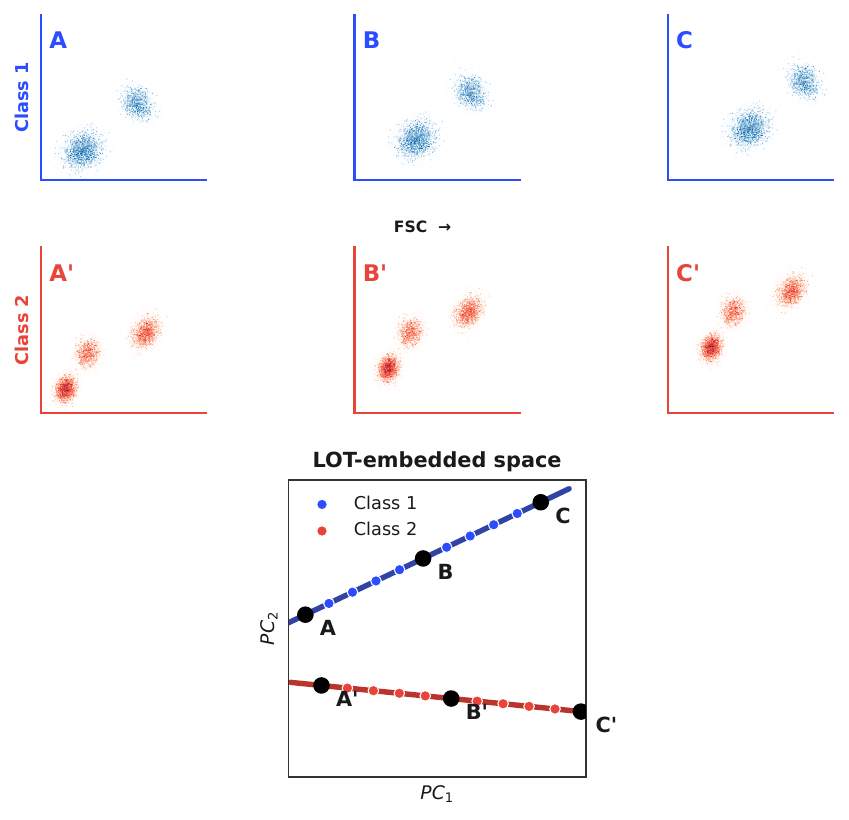}
    \caption{Visualization of the LOT embedding geometry using Principal Component Analysis, illustrating how the data geometry changes across different template classes. Selected samples from synthetic class 1 (blue) and class 2 (red) are inverted to demonstrate translational shifts in the underlying distribution/cytometry space.}
    \label{template_model}
\end{figure}

For ease of visualization, we first consider a single deformation type: translation of the template along a random axis, as shown in Fig.~\ref{template_model}. Each resulting distribution is embedded as a fixed-length vector that encodes its high-dimensional distributional structure. We then project these embeddings into a two-dimensional principal-component space, where each point corresponds to one full distribution. Thus, distances and separations in this space reflect differences between distributions, rather than differences between individual cells or measurements.

Figure~\ref{template_model} shows that the two classes form distinct linear manifolds in the projected embedding space, yielding strict linear separability in this controlled example. Consequently, a standard linear classifier achieves perfect discrimination. Importantly, this behavior is not restricted to simple translational deformations of the underlying distributions. Under suitable admissibility conditions, broader classes of geometric and structural deformations induce class-specific linear subspaces that remain separable under the proposed embedding. A rigorous mathematical characterization of these deformation models and the corresponding conditions for linear separability is developed in \cite{pointcloud3d,aldroubi2021partitioning,moosmuller2023linear}. 

This synthetic proof of concept illustrates how the proposed embedding captures class-specific distributional geometry. We next extend the analysis to real-world flow cytometry data, evaluating the model’s ability to distinguish AML samples from normal samples.

\subsubsection*{Acute myeloid leukemia (AML) classification}
\paragraph{Experimental Setup.}
To evaluate the effectiveness of the proposed \emph{Linearized Optimal Transport} (LOT) embedding for flow-cytometry measurements from real patient cohorts, we assessed diagnostic classification performance on two benchmark datasets. For each dataset, patient samples were embedded into the LOT space using a dataset-specific reference patient following the framework introduced in \cite{lot_main,pointcloud3d}. The optimal transport maps required for the embedding were computed using linear programming algorithms implemented in the POT library \cite{flamary2021pot,flamary2026pot}. Classification was subsequently performed using simple downstream classifiers operating on the resulting LOT representations.

We compared the proposed approach against several established methods for set-structured cytometry analysis. These included AttentionMIL \cite{ilse2018attention}, an attention-based multiple-instance learning framework that assigns diagnostic relevance weights to individual cells; CytoSet \cite{yi2021cytoset}, a permutation-invariant architecture derived from DeepSets \cite{deepsets} that aggregates cell-level embeddings into sample-level representations; PointNet++ \cite{pointnet} and DGCNN \cite{dgcnn}, two widely used deep-learning architectures for point-cloud classification; CellCNN \cite{arvaniti2017_cellcnn}, a convolutional framework specifically designed for identifying disease-associated cell populations; and FlowSOM \cite{bucklein2019flowsom}, a clustering-based representation baseline that summarizes cellular populations through self-organizing maps. Within the LOT embedding framework, we considered several classical classifiers, including logistic regression (LR), support vector machines (SVM), random forests (RF), extremely randomized trees (ET), and the proposed \emph{Nearest Subspace Classifier} (NSC).

All methods were evaluated on high-dimensional flow-cytometry datasets derived from the BLAST-110 acute myeloid leukemia (AML) study \cite{BLAST110_DATA} and the FlowCAP-II AML challenge \cite{AML_DATA}. We formulated binary diagnostic classification tasks in which each patient was represented by multiple diagnostic panels, with each panel containing either $500$, $1{,}000$, or $5{,}000$ randomly sampled single-cell measurements. To assess robustness under limited data availability, a common constraint in clinical studies, we constructed few-shot learning settings by varying the number of labeled training samples per class ($k$-shots). Performance was evaluated across ten independent stratified train--test splits and reported as the mean and standard deviation across repetitions.

\paragraph{Balanced Accuracy.}
Because the diagnostic tasks are not perfectly class-balanced, we report \emph{balanced accuracy} (BA) rather than conventional accuracy. Balanced accuracy is defined as the mean of the per-class recalls,
\begin{equation}
    \mathrm{BA}
    :=
    \frac{1}{C}
    \sum_{c=1}^{C}
    \frac{\mathrm{TP}_c}
         {\mathrm{TP}_c + \mathrm{FN}_c},
    \label{eq:balanced_accuracy}
\end{equation}
where $C$ denotes the number of classes and $\mathrm{TP}_c$ and $\mathrm{FN}_c$ represent the true positives and false negatives for class $c$, respectively. For binary classification problems, this reduces to
\[
\mathrm{BA}
=
\frac{1}{2}
\left(
\mathrm{sensitivity}
+
\mathrm{specificity}
\right).
\]

Unlike raw accuracy, which can be artificially inflated by always predicting the majority class, balanced accuracy assigns equal importance to each class and therefore provides a more reliable measure of discriminative performance under class imbalance. This property is particularly important in clinical applications, where the diagnostically relevant class is often the minority population.

Patient-level predictions were obtained through \emph{Score Fusion} (SF), whereby class probabilities from individual panels were averaged prior to final label assignment. This aggregation strategy preserves soft confidence information across panels while producing a unified patient-level prediction.

\begin{table*}[h]
\centering
\caption{\textbf{Multi-Tube Diagnostic Classification Results:  Balanced Accuracy (\%) across all methods, cell counts ($N$), and few-shot regimes ($k$).} Performance reported as $\text{Mean} \pm \text{Standard Deviation}$ across 10 independent stratified train/test splits. Within each split, a balanced few-shot training subset samples (k) per class was randomly selected from the training set, while evaluation was performed on the corresponding held-out test set. Multi-tube (\texttt{score\_fusion} is applied across all diagnostic panels. Bold font denotes the winning classifier for each experimental condition (including ties).}
\label{tab:score_fusion_scaling}
\small
\setlength{\tabcolsep}{4.5pt}
\renewcommand{\arraystretch}{1.0}
\resizebox{\textwidth}{!}{%
\begin{tabular}{ll cccc cccc}
\toprule
\multirow{2}{*}{\textbf{$N$}} & \multirow{2}{*}{\textbf{Method}} & \multicolumn{4}{c}{\textbf{BLAST110 (Binary AML vs. NBM)}} & \multicolumn{4}{c}{\textbf{FLOWCAP-II (Diagnostic AML vs. Normal)}} \\
\cmidrule(lr){3-6} \cmidrule(lr){7-10}
 & & $\mathbf{k=2}$ & $\mathbf{k=4}$ & $\mathbf{k=6}$ & $\mathbf{k=8}$ & $\mathbf{k=4}$ & $\mathbf{k=8}$ & $\mathbf{k=12}$ & $\mathbf{k=16}$ \\
\midrule
\multirow{11}{*}{\rotatebox{90}{$\mathbf{500}$}} & LOT +  NSC & $91.2 \pm 7.5$ & $95.7 \pm 3.6$ & $95.8 \pm 3.8$ & $97.0 \pm 1.9$ & $87.5 \pm 5.2$ & $90.8 \pm 4.8$ & $\mathbf{92.7 \pm 2.5}$ & $\mathbf{94.3 \pm 1.9}$ \\
 & LOT +  LR & $\mathbf{92.3 \pm 9.0}$ & $\mathbf{96.3 \pm 2.9}$ & $96.3 \pm 2.9$ & $97.7 \pm 1.6$ & $85.9 \pm 4.8$ & $90.1 \pm 5.5$ & $91.4 \pm 3.2$ & $92.2 \pm 1.8$ \\
 & LOT +  SVM & $61.0 \pm 40.6$ & $93.7 \pm 4.7$ & $95.7 \pm 3.6$ & $\mathbf{98.0 \pm 1.7}$ & $86.8 \pm 5.5$ & $91.1 \pm 4.2$ & $92.1 \pm 2.3$ & $93.5 \pm 1.8$ \\
 & LOT +  RF & $85.8 \pm 9.2$ & $95.2 \pm 2.5$ & $95.0 \pm 2.4$ & $95.8 \pm 3.3$ & $88.3 \pm 4.9$ & $90.0 \pm 5.5$ & $92.0 \pm 1.5$ & $92.8 \pm 1.8$ \\
 & LOT +  ET & $89.7 \pm 6.9$ & $94.5 \pm 3.5$ & $95.7 \pm 2.2$ & $96.2 \pm 3.1$ & $89.0 \pm 4.7$ & $\mathbf{91.3 \pm 2.8}$ & $92.3 \pm 1.7$ & $93.0 \pm 1.7$ \\
 & DGCNN & $90.7 \pm 7.7$ & $93.8 \pm 5.2$ & $\mathbf{96.4 \pm 2.4}$ & $97.0 \pm 2.6$ & $88.8 \pm 4.3$ & $90.9 \pm 3.7$ & $92.1 \pm 2.2$ & $92.8 \pm 2.0$ \\
 & AttentionMIL  & $90.5 \pm 8.5$ & $92.6 \pm 5.3$ & $95.8 \pm 3.5$ & $95.4 \pm 3.6$ & $88.8 \pm 4.1$ & $90.9 \pm 4.1$ & $92.1 \pm 2.3$ & $92.3 \pm 1.9$ \\
 & CytoSet  & $89.4 \pm 8.7$ & $92.0 \pm 5.2$ & $95.2 \pm 3.7$ & $96.2 \pm 2.3$ & $\mathbf{89.7 \pm 3.6}$ & $90.7 \pm 4.5$ & $91.4 \pm 2.7$ & $92.3 \pm 1.7$ \\
 & PointNet++  & $83.4 \pm 13.2$ & $89.2 \pm 7.7$ & $92.5 \pm 5.1$ & $93.7 \pm 4.3$ & $82.3 \pm 8.5$ & $87.2 \pm 6.4$ & $88.8 \pm 4.7$ & $90.3 \pm 4.3$ \\
 & CellCNN  & $77.3 \pm 12.1$ & $84.4 \pm 7.5$ & $89.0 \pm 5.2$ & $90.5 \pm 5.2$ & $75.4 \pm 8.6$ & $81.2 \pm 6.8$ & $84.0 \pm 5.9$ & $85.6 \pm 4.4$ \\
 & FlowSOM  & $77.5 \pm 12.4$ & $85.5 \pm 7.2$ & $91.5 \pm 5.1$ & $91.1 \pm 3.8$ & $83.0 \pm 6.8$ & $88.1 \pm 4.9$ & $89.8 \pm 4.1$ & $92.1 \pm 3.0$ \\
\midrule
\multirow{11}{*}{\rotatebox{90}{$\mathbf{1000}$}} & LOT +  NSC & $89.8 \pm 7.1$ & $91.5 \pm 8.9$ & $94.2 \pm 6.1$ & $96.2 \pm 4.0$ & $87.5 \pm 5.2$ & $90.8 \pm 4.8$ & $\mathbf{92.7 \pm 2.5}$ & $\mathbf{94.3 \pm 1.9}$ \\
 & LOT +  LR & $\mathbf{92.0 \pm 6.0}$ & $\mathbf{97.0 \pm 2.9}$ & $95.3 \pm 4.8$ & $\mathbf{98.0 \pm 1.7}$ & $85.9 \pm 4.8$ & $90.1 \pm 5.5$ & $91.4 \pm 3.2$ & $92.2 \pm 1.8$ \\
 & LOT + SVM & $61.5 \pm 40.7$ & $96.2 \pm 3.2$ & $\mathbf{96.5 \pm 2.5}$ & $\mathbf{98.0 \pm 1.7}$ & $86.8 \pm 5.5$ & $91.1 \pm 4.2$ & $92.1 \pm 2.3$ & $93.5 \pm 1.8$ \\
 & LOT +  RF & $88.3 \pm 8.5$ & $94.7 \pm 1.7$ & $95.7 \pm 2.2$ & $95.5 \pm 2.9$ & $88.3 \pm 4.9$ & $90.0 \pm 5.5$ & $92.0 \pm 1.5$ & $92.8 \pm 1.8$ \\
 & LOT +  ET & $91.0 \pm 6.9$ & $95.2 \pm 3.4$ & $95.8 \pm 2.9$ & $95.8 \pm 3.6$ & $89.0 \pm 4.7$ & $\mathbf{91.3 \pm 2.8}$ & $92.3 \pm 1.7$ & $93.0 \pm 1.7$ \\
 & DGCNN  & $91.2 \pm 8.0$ & $93.6 \pm 4.8$ & $95.9 \pm 2.8$ & $97.0 \pm 2.2$ & $88.8 \pm 4.3$ & $90.9 \pm 3.7$ & $92.1 \pm 2.2$ & $92.8 \pm 2.0$ \\
 & AttentionMIL & $90.8 \pm 7.9$ & $92.5 \pm 5.1$ & $95.5 \pm 3.8$ & $95.4 \pm 3.4$ & $88.8 \pm 4.1$ & $90.9 \pm 4.1$ & $92.1 \pm 2.3$ & $92.3 \pm 1.9$ \\
 & CytoSet & $90.3 \pm 7.6$ & $92.3 \pm 5.2$ & $95.2 \pm 4.3$ & $96.5 \pm 1.9$ & $\mathbf{89.7 \pm 3.6}$ & $90.7 \pm 4.5$ & $91.4 \pm 2.7$ & $92.3 \pm 1.7$ \\
 & PointNet++ & $83.2 \pm 11.4$ & $88.5 \pm 8.5$ & $92.4 \pm 4.6$ & $93.3 \pm 4.5$ & $82.3 \pm 8.5$ & $87.2 \pm 6.4$ & $88.8 \pm 4.7$ & $90.3 \pm 4.3$ \\
 & CellCNN  & $79.2 \pm 10.6$ & $85.3 \pm 9.3$ & $89.5 \pm 5.1$ & $91.9 \pm 3.9$ & $75.4 \pm 8.6$ & $81.2 \pm 6.8$ & $84.0 \pm 5.9$ & $85.6 \pm 4.4$ \\
 & FlowSOM  & $77.0 \pm 13.0$ & $87.4 \pm 7.3$ & $90.2 \pm 4.4$ & $92.7 \pm 3.8$ & $83.0 \pm 6.8$ & $88.1 \pm 4.9$ & $89.8 \pm 4.1$ & $92.1 \pm 3.0$ \\
\midrule
\multirow{11}{*}{\rotatebox{90}{$\mathbf{5000}$}} & LOT +  NSC & $89.7 \pm 8.1$ & $96.0 \pm 3.7$ & $\mathbf{97.7 \pm 1.6}$ & $\mathbf{98.0 \pm 1.7}$ & $87.5 \pm 5.2$ & $90.8 \pm 4.8$ & $\mathbf{92.7 \pm 2.5}$ & $\mathbf{94.3 \pm 1.9}$ \\
 & LOT +  LR & $90.0 \pm 7.3$ & $96.7 \pm 3.0$ & $97.0 \pm 2.9$ & $\mathbf{98.0 \pm 1.7}$ & $85.9 \pm 4.8$ & $90.1 \pm 5.5$ & $91.4 \pm 3.2$ & $92.2 \pm 1.8$ \\
 & LOT +  SVM & $28.7 \pm 24.6$ & $90.0 \pm 6.1$ & $95.7 \pm 3.2$ & $97.5 \pm 2.6$ & $86.8 \pm 5.5$ & $91.1 \pm 4.2$ & $92.1 \pm 2.3$ & $93.5 \pm 1.8$ \\
 & LOT +  RR & $89.0 \pm 8.6$ & $95.8 \pm 3.1$ & $97.3 \pm 2.1$ & $97.7 \pm 2.2$ & $88.3 \pm 4.9$ & $90.0 \pm 5.5$ & $92.0 \pm 1.5$ & $92.8 \pm 1.8$ \\
 & LOT +  ET & $90.0 \pm 8.9$ & $\mathbf{97.0 \pm 1.9}$ & $97.0 \pm 1.9$ & $97.7 \pm 1.6$ & $89.0 \pm 4.7$ & $\mathbf{91.3 \pm 2.8}$ & $92.3 \pm 1.7$ & $93.0 \pm 1.7$ \\
 & DGCNN  & $90.3 \pm 7.8$ & $93.0 \pm 5.5$ & $95.4 \pm 3.5$ & $96.5 \pm 2.7$ & $88.8 \pm 4.3$ & $90.9 \pm 3.7$ & $92.1 \pm 2.2$ & $92.8 \pm 2.0$ \\
 & AttentionMIL& $\mathbf{91.6 \pm 7.7}$ & $92.6 \pm 5.6$ & $95.1 \pm 4.3$ & $95.3 \pm 3.5$ & $88.8 \pm 4.1$ & $90.9 \pm 4.1$ & $92.1 \pm 2.3$ & $92.3 \pm 1.9$ \\
 & CytoSet  & $91.2 \pm 7.5$ & $91.8 \pm 5.2$ & $94.2 \pm 4.7$ & $96.2 \pm 2.3$ & $\mathbf{89.7 \pm 3.6}$ & $90.7 \pm 4.5$ & $91.4 \pm 2.7$ & $92.3 \pm 1.7$ \\
 & PointNet++  & $69.1 \pm 15.0$ & $84.6 \pm 11.3$ & $89.7 \pm 6.0$ & $90.5 \pm 4.8$ & $82.3 \pm 8.5$ & $87.2 \pm 6.4$ & $88.8 \pm 4.7$ & $90.3 \pm 4.3$ \\
 & CellCNN  & $80.7 \pm 10.9$ & $86.9 \pm 7.7$ & $90.8 \pm 4.6$ & $92.2 \pm 3.6$ & $75.4 \pm 8.6$ & $81.2 \pm 6.8$ & $84.0 \pm 5.9$ & $85.6 \pm 4.4$ \\
 & FlowSOM  & $81.9 \pm 12.3$ & $89.0 \pm 6.7$ & $93.6 \pm 3.9$ & $94.4 \pm 3.0$ & $83.0 \pm 6.8$ & $88.1 \pm 4.9$ & $89.8 \pm 4.1$ & $92.1 \pm 3.0$ \\
\bottomrule
\end{tabular}
}
\vspace{5pt}

\end{table*}
\paragraph{Classification Performance.} Table~\ref{tab:score_fusion_scaling} summarizes the mean balanced accuracy ($\pm$ standard deviation) across ten independent train--test splits for all methods, cell counts, and few-shot settings. Overall, classifiers operating on LOT embeddings achieved performance comparable to or exceeding substantially more complex deep-learning baselines across both datasets. Importantly, strong results were obtained with multiple downstream classifiers, suggesting that a substantial portion of the discriminative power arises from the LOT representation itself rather than from a particular classification model. 

On the FlowCAP-II benchmark, LOT-based methods consistently achieved competitive performance across all few-shot regimes. In particular, LOT+NSC attained balanced accuracies of $92.7\%$ and $94.3\%$ for $k=12$ and $k=16$, respectively. These results were obtained without iterative optimization, architecture-specific design choices, or extensive hyperparameter tuning. Other classifiers operating on LOT embeddings, including logistic regression, support vector machines, random forests, and extremely randomized trees, also demonstrated strong and stable performance, indicating that the LOT representation captures disease-relevant information in a form that can be effectively exploited by a range of learning algorithms. 

The advantages of LOT embeddings are similarly evident on the BLAST110 dataset. Across most experimental settings, classifiers built on LOT representations rank among the strongest-performing approaches and frequently outperform specialized deep-learning architectures such as AttentionMIL, CytoSet, DGCNN, PointNet++, and CellCNN. The proposed NSC achieves particularly competitive performance in low-shot settings, where limited training data often hinder the effectiveness of deep neural networks. More broadly, the consistently strong performance observed across several classical classifiers suggests that the LOT embedding provides a robust and informative representation of patient-specific cellular distributions. 

An additional observation is that increasing the number of sampled cells from $500$ to $1{,}000$ generally improves predictive performance and reduces variance for most methods. However, further increasing the number of cells to $5{,}000$ yields only modest gains despite increased computational cost. These results suggest that approximately $1{,}000$ cells per panel provide a favorable trade-off between classification accuracy, robustness, and computational efficiency. Interestingly, classification performance on the FlowCAP-II benchmark remains relatively stable across cellular subsampling depths, indicating that the key disease-discriminative information can be captured from comparatively small subsets of cells. 

Beyond predictive accuracy, the LOT-based framework offers several practical advantages. Deep-learning approaches typically require iterative optimization, hyperparameter tuning, and GPU resources, whereas LOT+NSC combines a deterministic embedding stage with a parameter-free geometric classification rule. Once the LOT embedding has been computed, classification reduces to evaluating distances to class-specific subspaces, resulting in negligible training and inference overhead. Consequently, the proposed framework provides a favorable combination of predictive performance, interpretability, reproducibility, and computational efficiency. 

Taken together, these results demonstrate that LOT embeddings provide a highly informative representation of patient-specific cellular distributions, enabling simple classical classifiers to perform on par with, and often better than, substantially more complex deep-learning models. Among the evaluated approaches, LOT+NSC offers an attractive balance of accuracy, robustness, and efficiency while requiring no architecture-specific tuning. These properties make the proposed framework particularly well suited for clinical flow-cytometry applications, where labeled cohorts are limited and reproducible analytical pipelines are essential.

\subsection*{Technical variability, normalization challenges, and robustness of the proposed framework}

Flow cytometry data are affected by systematic scaling, translation, and batch-related artifacts that hinder reliable cross-sample and cross-experiment comparisons. Fluorescence intensities often span several orders of magnitude, producing extreme dynamic ranges that complicate visualization and statistical modeling. In addition, fluorescence compensation and background correction can introduce zero or negative values, limiting the suitability of standard logarithmic transformations and motivating specialized transformations such as logicle, biexponential, or Box--Cox-based approaches \cite{lo2009flowclust}. 

Biologically comparable cell populations may also shift across samples because of instrument variability, detector gain, staining efficiency, antibody lot differences, compensation errors, and batch effects, resulting in systematic misalignment of marker intensities \cite{finak2014high}. These technical artifacts can propagate to downstream analyses, leading to inconsistent gating, unstable clustering or dimensionality reduction, and reduced generalization of machine-learning models in high-dimensional cytometry data \cite{saeys2016computational}.

To mitigate these effects, a range of normalization strategies has been proposed, including nonlinear transformations for variance stabilization (e.g., hyberbolic arcsine, biexponential), bead-based calibration, distribution-based methods such as quantile normalization and landmark alignment, and batch-correction algorithms including ComBat and CytoNorm \cite{van2020cytonorm}. However, many approaches rely on strong assumptions of distributional similarity and risk over-correcting data or removing genuine biological variation. In particular, marker-specific behaviors and rare populations remain difficult to normalize consistently across heterogeneous experiments, frequently requiring substantial expert intervention.

The proposed optimal transport-based embedding, combined with subspace-based classification, is designed to mitigate common sources of technical variability in flow cytometry data. In particular, the embedding represents each sample through its transport with respect to a fixed reference distribution. Under translation or scaling of the original distribution, the corresponding embedded representation changes by a structured transformation, such as an additive constant or global scaling factor, rather than by an arbitrary distortion of the sample geometry. Related properties of optimal transport maps have been discussed in \cite{zhuang2025local}.

This structure helps preserve the relative ordering and alignment of cellular measurements with respect to the reference distribution, even when absolute marker intensities shift across experiments. As a result, biologically comparable cell populations remain more consistently aligned in the embedded representation under systematic scaling or translation artifacts. The subspace-based classifier further exploits this structure because classification depends on the geometry of the embedded samples rather than on absolute marker intensities alone. Since subspace-based classification is insensitive to global scaling and constant shifts in the embedded feature space, the combined embedding-classification framework can reduce the impact of these technical artifacts.

The framework also remains stable under mild distributional perturbations, such as small deviations from symmetry or modest skewness, provided that these perturbations do not substantially alter the underlying class-specific geometry. Consequently, samples from the same biological class that differ because of technical scaling, translation, or mild distributional distortion can still occupy a coherent region of the embedded space. This preservation of class geometry helps explain why only a small number of training samples may be sufficient to characterize each class subspace, supporting strong classification performance even when training data are limited.

\subsection*{Visualization of class-specific features in AML}

The proposed embedding-based framework enables visualization and quantitative analysis of class-specific marker variation. In the learned embedding space, we compute the class means for the two populations using the training data and define a principal axis as the direction connecting these mean vectors. This axis represents the dominant mode of separation between classes and captures systematic differences in marker-expression patterns.

Both training and held-out test samples are projected onto this axis. In both sets, the two classes are well separated, indicating that the embedding generalizes beyond the training data and captures reproducible class-associated differences. Importantly, this one-dimensional axis provides an interpretable coordinate along which continuous variation between normal bone marrow and acute myeloid leukemia (AML) samples can be examined.

\begin{figure}[htb!]
    \centering
    \includegraphics[width=0.8\textwidth]{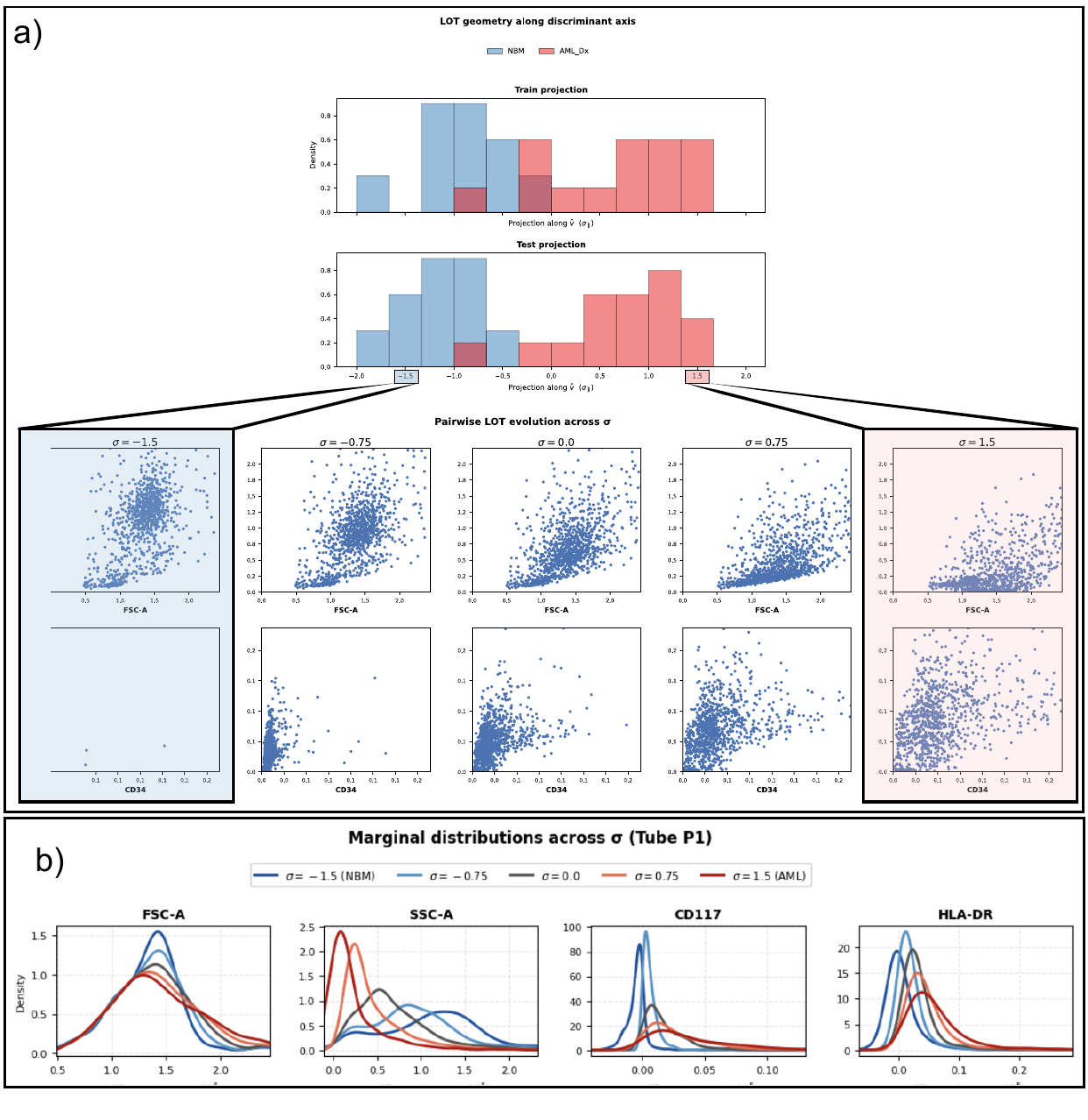}
    \caption{\textbf{Visualization of class-specific feature variation by analyzing trajectories in embedding space and mapping them back to the original feature space.}\textbf{(a)} Projection of samples onto the principal discriminative axis, defined as the direction connecting the mean embedding vectors of the two classes computed from the training data. Training samples are shown on the left and held-out test samples on the right. In both sets, the two classes separate clearly along this axis, indicating that the learned embedding captures a robust and generalizable mode of class-specific variation.\textbf{(b)} Feature-distribution changes along the principal axis. The axis is divided into intervals spanning $-1.5\sigma$ to $+1.5\sigma$, where $\sigma$ denotes the standard deviation of the sample projections along the axis. Six pairwise feature-distribution panels are shown, each containing distributions corresponding to different locations along the axis. This representation shows how marker-expression patterns vary systematically along the class-separating direction, supporting validation of known phenotypic trends and identification of candidate class-associated patterns.}
    \label{fig:combined}
\end{figure}

To characterize how cytometry features change along this axis, we analyze samples at different locations along the projection. At selected intervals, we visualize pairwise marker distributions (Figure~\ref{fig:combined}) and compare the observed trends with established immunophenotypic patterns of AML blast populations. In this dataset, AML-associated regions show reduced side scatter (SSC), increased forward scatter (FSC), dimmer CD5 expression, and increased expression of markers such as CD34, CD117, and HLA-DR. These findings are broadly consistent with immature myeloid phenotypes observed in many AML cases, although AML immunophenotypes are heterogeneous and marker expression can vary across subtypes.

Beyond recapitulating established phenotypic patterns, this framework can reveal additional class-associated trends by examining feature evolution along the embedding axis. The approach also supports data-driven hypothesis generation, allowing candidate marker associations to be further evaluated and validated in independent datasets.

\subsection*{Estimation of Blast Percentage in White Blood Cells for Residual Disease Monitoring}

Measurable residual disease (MRD) in AML refers to residual leukemic cells that persist after therapy below the limit of conventional morphologic detection. In flow-cytometric MRD assessment, disease burden is commonly quantified as the proportion of aberrant leukemic blasts among total white blood cells (WBCs). This measurement is an established biomarker associated with relapse risk and clinical outcome in AML. In current clinical practice, MRD is frequently assessed using multiparameter flow cytometry through the identification of leukemia-associated immunophenotypes (LAIPs) by expert-driven manual gating. However, this process is labor-intensive, subjective, and sensitive to small gating differences, particularly near the clinically relevant MRD positivity threshold of $0.1\%$, motivating the development of more objective computational approaches for MRD assessment (cMRD) \cite{BLAST110_DATA}.

\begin{figure}[H]
  \centering
  \includegraphics[width=0.9\textwidth]{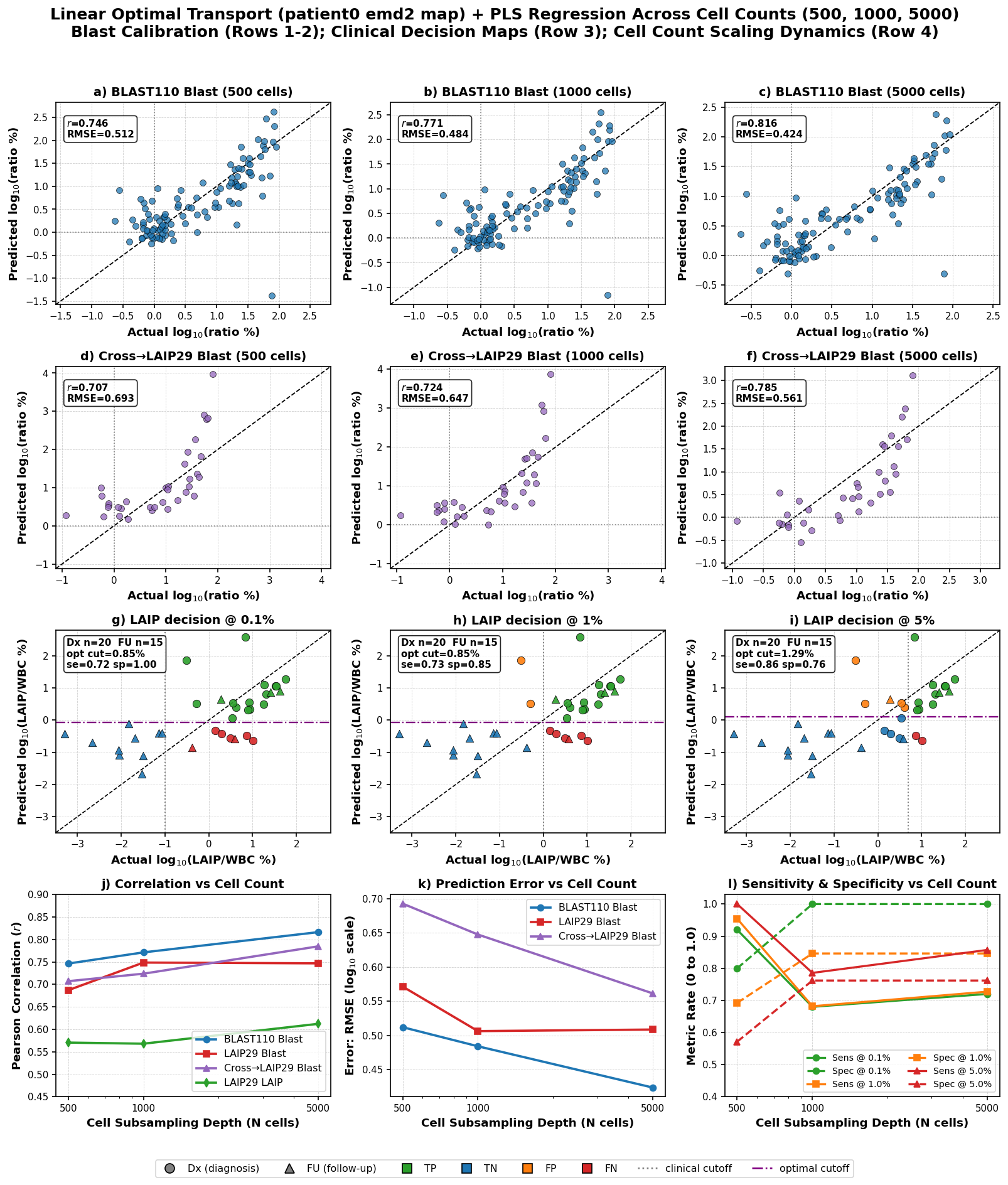}
  \caption{\textbf{Partial Least Squares (PLS) regression for leukemic blast and LAIP quantification across varying cell subsampling depths.} \textbf{(a--f)} Blast calibration and cross-cohort prediction show improved Pearson correlation ($r$) and reduced error (RMSE) as sampled cells increase from 500 to 5000. \textbf{(g--i)} Clinical decision maps at 5000 cells evaluating LAIP diagnostic utility at 0.1\%, 1.0\%, and 5.0\% thresholds for Diagnosis (Dx, $n=20$) and Follow-up (FU, $n=15$) samples, utilizing an optimal cutoff to balance sensitivity (se) and specificity (sp). \textbf{(j--l)} Cell count scaling dynamics indicate that performance metrics improve and stabilize  with increase number of cell counts.}
  
  \label{fig:MRD_results2}
\end{figure}

Direct quantification of disease burden from raw flow-cytometry measurements remains challenging because each sample consists of an unordered collection of thousands of cells in a high-dimensional marker space, with no natural correspondence between cells across patients. Consequently, most computational approaches rely on gating, clustering, or hand-crafted summary features prior to downstream analysis. The LOT framework provides an alternative representation by aligning each sample to a common reference distribution and mapping it into a fixed-length Euclidean vector representation. This transformation enables the application of conventional statistical learning methods directly to sample-level cellular distributions. 

To investigate whether LOT embeddings capture information relevant to disease burden, we modeled blast abundance as a continuous outcome using partial least squares (PLS) regression applied directly to the LOT embeddings. The regression model was trained to estimate the blast-to-WBC percentage on a $\log_{10}$ scale, $\log_{10}(\mathrm{Blast/WBC}\,\%)$, from paired flow-cytometry measurements and blast annotations. For this analysis, LOT embeddings were generated using Tube~1 measurements only. This design choice reflects the incomplete availability of additional tubes for several samples in the LAIP29 dataset and provides an initial evaluation of continuous disease quantification within the LOT framework. All predictive evaluations were performed using 5-fold cross-validation with 20 independent randomized repetitions to obtain stable out-of-sample performance estimates. 

Figure~\ref{fig:MRD_results2} summarizes regression performance across cellular subsampling depths ranging from $N=500$ to $N=5000$ cells. On the BLAST110 cohort ($n=107$), performance improved as the number of sampled cells increased, yielding a Pearson correlation of $r=0.816$ at $N=5000$. Similarly, when trained on the BLAST110 cohort and evaluated on the independent LAIP29 dataset ($n=35$), the model achieved a correlation of $r=0.785$ at $N=5000$. These results indicate that LOT embeddings preserve population-level information associated with leukemic blast burden and support generalization across independently acquired cohorts. The observed improvement with increasing cellular sampling depth further suggests that richer representations of the underlying cellular distribution may benefit quantitative disease estimation. 

Beyond continuous blast quantification, we evaluated the clinically relevant task of MRD detection using LAIP thresholds of $0.1\%$, $1\%$, and $5\%$. The LAIP29 cohort ($n=35$) contains 20 diagnosis (Dx) samples and 15 follow-up (FU) samples. Consistent with clinical expectations, diagnosis samples exhibited substantially higher leukemic burden (median $\mathrm{LAIP/WBC}=7.6\%$), whereas follow-up samples primarily occupied the low-burden residual-disease regime (median $0.03\%$). To balance sensitivity and specificity, decision thresholds were selected by maximizing Youden's index, \[ J = \mathrm{sensitivity} + \mathrm{specificity} - 1. \] 

At the corresponding operating point using $N=5000$ cells, the model detected MRD positivity at the clinically important $0.1\%$ threshold with $72\%$ sensitivity and $100\%$ specificity ($J=0.72$). Balanced performance was similarly observed at the $1\%$ and $5\%$ thresholds, yielding sensitivities of $0.73$ and $0.86$ with corresponding specificities of $0.85$ and $0.76$, respectively. 

Taken together, these findings demonstrate that LOT embeddings capture biologically meaningful variation associated with leukemic blast burden and support quantitative estimation of continuous MRD-related outcomes directly from high-dimensional flow-cytometry distributions. By mapping sample-level cellular measurements into a common vector representation, the proposed framework enables computational MRD assessment without explicit gating, predefined leukemia-associated phenotypes, or marker-specific feature engineering. While additional validation on larger cohorts will be important, these results suggest that LOT-based regression provides a promising and scalable framework for objective MRD monitoring in AML.

\subsection*{Limitations and future work}
To reduce the computational cost of optimal transport estimation, our current implementation constructs LOT embeddings using randomly subsampled cellular populations containing $N=500$, $1{,}000$, or $5{,}000$ cells per sample. While the experimental results demonstrate that accurate and robust classification can be achieved with relatively small subsets of cells, subsampling may not fully capture extremely rare cellular populations or subtle distributional structures present in the complete specimen. Although performance generally improves as the number of sampled cells increases, the gains beyond $1{,}000$ cells are modest for most classification tasks. Future advances in scalable optimal transport algorithms may enable the efficient use of substantially larger cellular populations, further improving the fidelity of the resulting embeddings while preserving computational tractability. 

A second limitation concerns the treatment of multi-tube flow-cytometry measurements. In the current framework, each tube is analyzed independently across patients, and patient-level predictions are obtained by aggregating tube-specific outputs through score fusion. This strategy is necessitated by the availability of labels only at the patient level and has been shown to perform well in practice. However, individual tubes may contain markers with varying relevance to a particular disease phenotype, and treating all tubes independently may limit the ability to exploit complementary information distributed across the full marker panel. Future work will focus on developing unified representations that integrate information across all tubes simultaneously, enabling joint modeling of the complete immunophenotypic profile. Such an approach could improve the detection of disease-associated cellular patterns, enhance prediction accuracy, and provide a more comprehensive characterization of patient-specific immune and leukemic phenotypes.

\section*{Discussion} 
We present \textbf{\textit{FlowLOT}}, a distribution-level framework for flow cytometry analysis that leverages linearized optimal transport to enable principled comparison, classification, and interpretation of high-dimensional cellular populations. By representing each cytometry sample as an empirical probability distribution and embedding these distributions into a vector space that preserves their geometric structure, \textbf{\textit{FlowLOT}} addresses long-standing challenges associated with permutation invariance, high dimensionality, and limited labeled data in cytometry studies. Our results demonstrate that this approach unifies classification, visualization, and measurable residual disease (MRD) inference within a single, mathematically transparent framework.

A central conceptual contribution of this work is the shift from cell-level or cluster-centric representations toward population-level geometric modeling. Conventional pipelines rely on intermediate steps such as manual gating, clustering, or parametric mixture modeling, each of which introduces additional assumptions and potential sources of variability. In contrast, \textbf{\textit{FlowLOT}} operates directly on empirical distributions, avoiding explicit cell-to-cluster assignments and reducing sensitivity to initialization, model order selection, or heuristic thresholds. This distribution-aware perspective is particularly well suited to clinical contexts such as acute myeloid leukemia, where disease-associated signals often manifest as subtle, coordinated shifts across multiple markers rather than as distinct, well-separated subpopulations.

The use of linearized optimal transport is key to achieving both computational tractability and interpretability. While full optimal transport distances are computationally expensive and do not naturally admit linear operations, the linearized embedding employed here enables closed-form, non-iterative classification under a generative deformation model. This property allows \textbf{\textit{FlowLOT}} to perform well even in low-sample regimes, which are common in clinical and translational studies. Importantly, the approximate linear structure induced by the embedding also clarifies why simple linear classifiers can succeed when class-specific variability follows structured geometric patterns, providing insight into the conditions under which the method is expected to generalize.

Beyond classification, the invertibility of the transport-based embedding enables direct visualization of differences between phenotypic classes in the original marker space. Unlike dimensionality reduction techniques such as UMAP or t-SNE, which may distort inter-marker relationships and lack a well-defined inverse, \textbf{\textit{FlowLOT}} supports biologically interpretable visualization of class-associated distributional shifts. This capability facilitates hypothesis generation and may aid domain experts in identifying marker combinations or population-level trends associated with disease progression or treatment response.

\textbf{\textit{FlowLOT}} also offers a unified approach to MRD estimation. Existing MRD pipelines typically depend on reference comparisons, clustering heuristics, and manual review, which can be sensitive to rare populations and overlapping phenotypes. By modeling patient samples and controls within the same distributional embedding and quantifying deviations in a continuous, geometry-aware manner, \textbf{\textit{FlowLOT}} provides a principled alternative that does not require explicit leukemic cluster identification. The strong correlation observed between \textbf{\textit{FlowLOT}}-derived estimates and clinically assessed MRD underscores the potential of population-level optimal transport metrics for sensitive disease monitoring.

Future work will focus on extending the framework in several directions. Incorporating explicit modeling of batch effects and technical variability within the transport formulation may further improve robustness across multicenter studies. Adapting the method to longitudinal data could enable direct modeling of disease trajectories as paths in distribution space. Finally, while this study focuses on flow cytometry, the underlying methodology is broadly applicable to other high-dimensional, point-cloud–like biological data modalities, including mass cytometry and single-cell proteomics.

In summary, \textbf{\textit{FlowLOT}} introduces a mathematically grounded, distribution-level approach to cytometry analysis that complements existing tools while addressing key limitations related to interpretability, sample efficiency, and population-level inference. By bridging optimal transport theory with practical cytometry applications, this framework provides a flexible foundation for future methodological developments in high-dimensional single-cell analysis.

\section*{Acknowledgements}
This work was funded in part by NIH awards GM130825, CA274499, and ONR award N000142212505.

\section*{Code availability}
Code is available at \url{https://github.com/naqibUVa/FlowLOT_v1}.
\section*{Data availability}
The flow-cytometry datasets used in this study are publicly available. The \textbf{BLAST110} and \textbf{LAIP29} datasets can be accessed through the Zenodo repository at \href{https://zenodo.org/records/11046402}{https://zenodo.org/records/11046402}. The \textbf{FlowCAP-II} AML dataset is available through FlowRepository at \href{http://flowrepository.org/id/FR-FCM-ZZYA}{http://flowrepository.org/id/FR-FCM-ZZYA}.
\bibliography{refs}

\section*{Mathematical Representation of \textit{FlowLOT}}

In this section, we present the mathematical formulation of the FlowLOT framework. We introduce a principled representation of multiparameter flow cytometry measurements, an analytic modeling strategy for class‑level variation, and an optimal transport–based embedding designed to address key analytical challenges in this field based on this formulation, we state explicit problem definitions and derive analytical solution approaches for classification, visualization, and quantification tasks in flow cytometry.

\subsection*{Component 1: Representation of flow cytometry (FCM) measurements as empirical probability distributions}

In multiparameter FCM, each cell is characterized by fluorescence intensity measurements from antibody-conjugated markers, together with light-scatter parameters such as forward scatter and side scatter. These measurements can be organized into a vector, so that each cell is represented as a point in a high-dimensional feature space, with each dimension corresponding to one measured parameter. A biological sample, consisting of many such cells, can therefore be viewed as a point cloud in this space. The collection of cells from a sample naturally defines an empirical probability distribution over the feature space. Specifically, a sample containing $N$ cells measured over $L$ markers can be represented as the discrete probability measure
\[
\mu = \frac{1}{N}\sum_{i=1}^{N}\delta_{x_i}, \qquad x_i \in \mathbb{R}^L,
\]
where $x_i$ denotes the multiparameter measurement vector of the $i$-th cell and $\delta_{x_i}$ is a Dirac measure centered at $x_i$. Here, $L$ denotes the number of measured channels (markers), which is dataset-specific due to variations in cytometry panels; a detailed listing of channels for each dataset is provided in the Materials and Datasets section.

\subsection*{Component 2: Analytical Modeling Perspective}

To model population-level variability across samples, we represent observed flow cytometry distributions as structured deformations of a representative population. These deformations are formalized using the \emph{push-forward} operation, denoted by $\sharp$. Intuitively, applying a push-forward means transporting every point in a distribution according to a transformation and examining the resulting distribution after this transformation. In the context of flow cytometry, this provides a way to describe coordinated shifts in marker-expression and light-scatter profiles across a cell population, rather than treating variability as independent cell-level noise.

Using this framework, we assume that samples belonging to a phenotypic class $k$ arise from a class-specific template distribution $\varphi^{(k)}$, where the superscript $(k)$ denotes the class label. Individual samples within the class are generated by smooth, invertible transformations $g_j$ drawn from a diffeomorphism group $\mathcal{G}$:
\begin{equation}
\mathcal{S}^{(k)}
=
\left\{
\mu^{(k)}_j \;\big|\;
\mu^{(k)}_j
=
{g_j}_\sharp\,\varphi^{(k)},
\;\;
g_j \in \mathcal{G}
\right\}.
\end{equation}

This analytical model provides an effective and compact mathematical description of population variability, treating biological differences as structured deformations of a common cellular landscape. By explicitly modeling variability in this way, the framework enables systematic, mathematically grounded analysis of flow cytometry data and directly supports downstream tasks such as classification, visualization, and quantitative inference.

\subsection*{Component 3: Embedding}

A key challenge in population-level flow cytometry analysis is that each sample is represented by an unordered set of single-cell measurements. In the empirical distribution representation
\[
\mu = \frac{1}{N}\sum_{i=1}^{N}\delta_{x_i}, \quad x_i \in \mathbb{R}^L,
\]
the indexing of cells carries no biological meaning, since the distribution is unchanged under any permutation of $\{x_i\}_{i=1}^N$. Thus, each sample is fundamentally an \emph{unordered collection} of cells. This lack of intrinsic ordering makes direct comparison between samples nontrivial and limits the direct application of standard statistical learning methods, which typically require structured, fixed-length representations. To address this, we introduce an optimal-transport-based embedding that aligns each empirical distribution to a common reference and represents it through the corresponding transport structure, producing a \emph{structured, fixed-length vector representation}.

To achieve this, we compare two samples, a reference distribution $\sigma$ and a target distribution $\mu$, using \emph{optimal transport (OT)}. OT defines a map
\[
T : \mathbb{R}^L \to \mathbb{R}^L
\]
that rearranges the mass of $\sigma$ to match $\mu$ while preserving total mass, expressed as
\[
T_{\sharp}\sigma = \mu.
\]
Among all such rearrangements, OT selects the one that minimizes the total squared displacement,
\[
W_2^2(\sigma,\mu)
=
\min_{T:\, T_{\sharp}\sigma=\mu}
\int_{\mathbb{R}^L} \|x - T(x)\|^2 \, d\sigma(x).
\]
This formulation provides a biologically meaningful notion of distance, as it identifies the most efficient way to deform one population of cells into another, capturing coordinated shifts in marker expression rather than individual cell identities.

In the empirical setting with a finite number of cells, this optimal transport problem reduces to finding an optimal matching between the cells of $\sigma$ and $\mu$. Although each sample is originally unordered, this matching induces a consistent correspondence: each cell in the reference is paired with exactly one cell in the target. By fixing an arbitrary but consistent ordering of the reference cells,
\[
\sigma = \{x_1^{(\sigma)}, \dots, x_N^{(\sigma)}\},
\]
the transport map naturally assigns to each index $i$ a corresponding cell in the target,
\[
x_i^{(\sigma)} \longmapsto T(x_i^{(\sigma)}),
\]
thereby introducing a shared coordinate system across samples.

Using this alignment, we define the embedding of the sample $\mu$ relative to $\sigma$ as the collection of transported points,
\[
\widehat{\mu} =
\big( T(x_1^{(\sigma)}), \; T(x_2^{(\sigma)}), \; \dots, \; T(x_N^{(\sigma)}) \big),
\]
which can be viewed as an element of $\mathbb{R}^{N \times L}$ (or equivalently $\mathbb{R}^{NL}$ after vectorization). This construction converts each sample into a fixed-length object defined over a common ordered index set, while remaining invariant to the original permutation of cells. In this sense, the embedding transforms an unordered distribution into an ordered vector representation by leveraging the reference-induced alignment.

This framework, known as \emph{linearized optimal transport (LOT)}, represents each sample by the transport required to align a shared reference distribution with that sample. Rather than treating each sample as an unordered collection of cells, LOT describes how fixed support points of the reference distribution are displaced toward the target distribution. For a reference support point $x_i^{(\sigma)}$, the transport map assigns a corresponding location $T(x_i^{(\sigma)})$ in the target feature space, and the embedding is obtained by collecting these mapped locations across all reference points.

In this way, the transformation
\[
\text{unordered cells} \;\longrightarrow\; \text{reference-aligned representation} \;\longrightarrow\; \text{fixed-length vector}
\]
becomes concrete: each coordinate block of the embedding corresponds to a fixed reference index and records the location to which that reference point is transported. Because the same reference distribution is used for all samples, these coordinate blocks have consistent meaning across samples, making the resulting representations directly comparable. Differences between samples therefore reflect how the \emph{same reference support points} are displaced across populations, capturing structured, population-level changes in marker expression and light-scatter profiles.

In the LOT representation, these shifts take a simple and interpretable form. Since each sample is described by the displacement of the same reference cells, transformations in the original space translate into operations on these displacements. For example, a global shift (translation) in marker expression appears as an additive term in the embedding, while a uniform scaling of the population corresponds to a multiplicative effect. More generally, the embedding of the $j^{th}$ sample in class $k$ can be written as
\[
\widehat{\mu}_j^{(k)} = g_j \circ \widehat{\varphi}^{(k)},
\]
where $\widehat{\varphi}^{(k)}$ is the embedded template and $g_j$ describes how the reference cells are further displaced to generate that sample. Expressed in this common coordinate system, even complex nonlinear differences in the original distributions become simpler and are often well approximated by linear operations, leading to low-dimensional, structured variability that is directly amenable to standard statistical analysis.  In our experiments, for each tube in both datasets, we selected a fixed reference sample by choosing a single patient (e.g., the first patient) and defining the corresponding empirical distribution as the reference. This reference distribution was constructed using a fixed number ($N$) of cell-level features, randomly sampled without replacement, ensuring a consistent and unbiased representation across all analyses.

\subsection*{Component 4: Applications}

\subsubsection*{Materials and Datasets}

\paragraph{Dataset description.}
We evaluated the proposed framework on one simulated dataset and two real-world AML-related flow cytometry datasets. The simulated dataset consists of two classes generated from class-specific template distributions via controlled translations along a prescribed direction, following the analytical construction introduced in this work. This setting provides a controlled environment for probing the geometric structure of the embedding.

The first real-world dataset was obtained from the FlowCAP-II initiative, a community benchmark for distinguishing acute myeloid leukemia (AML) samples from healthy controls \cite{AML_DATA}. The cohort includes 359 individuals (316 healthy and 43 AML), with measurements acquired using eight distinct 7-color panels and provided in standard FCS format. We excluded tubes 1 and 8, as they correspond to control panels rather than informative measurement channels. Classification was performed independently on each of the remaining six tubes, and subject-level predictions were obtained by majority voting across tubes, assigning an AML label if at least half of the tubes yielded positive predictions. The dataset is publicly available through FlowRepository (FR-FCM-ZZYA).

The second real-world dataset, BLAST110, contains samples from 110 individuals (90 AML and 20 healthy controls) \cite{BLAST110_DATA}. The AML cohort comprises both diagnostic ($n=30$) and follow-up ($n=60$) samples, spanning a wide range of blast proportions and capturing substantial biological heterogeneity associated with disease progression. For both datasets, we randomly drew $N \in \{500, 1000, 5000\}$ cells without replacement from each tube. This standardized the representation across samples and controlled for variations in total cell yield.

\paragraph{Preprocessing strategy.}

In standard flow cytometry pipelines, extensive preprocessing is routinely applied prior to downstream analysis: compensation for spectral overlap, nonlinear intensity transformations (e.g., logicle or arcsinh), batch-effect correction, cross-sample normalization, noise filtering, removal of debris and doublets, and expert-driven gating to define cell populations. Although these steps can improve interpretability, they introduce substantial dependence on manual intervention, domain expertise, and dataset-specific heuristics, which limits reproducibility and scalability.

In this study, we deliberately refrain from any preprocessing beyond what the released datasets already provide; in particular, we apply no additional transformation, normalization, filtering, or gating. This design isolates the intrinsic robustness of the proposed framework under minimally processed conditions, while reducing operator bias and reliance on labor-intensive curation. Concretely, the robustness we establish in the Results concerns translation- and scaling-type artifacts: the LOT embedding transforms predictably when each cell's measurement vector $\mathbf{x} \in \mathbb{R}^{L}$ is mapped as $\mathbf{x} \mapsto A\mathbf{x} + \mathbf{b}$, where $A = \operatorname{diag}(a_1,\dots,a_L) \succ 0$ applies a per-marker (axis-aligned) scaling and $\mathbf{b} \in \mathbb{R}^{L}$ is a per-marker offset. This axis-aligned, commutative family captures many of the per-marker gain and baseline discrepancies that motivate normalization and rescaling steps.

Spectral unmixing (and, in conventional cytometry, compensation) is a categorically different operation and should not be conflated with these artifacts. It is not a correction applied within a fixed feature space but a change of representation between spaces \cite{Novo2013, Rajwa2025}. The measurement model is linear: the raw detector intensities relate to the underlying marker abundances through a mixing matrix $\mathbf{M}$ whose entries encode each marker's spectral signature across detectors, and unmixing recovers the abundances by applying $\mathbf{M}^{-1}$ (or its pseudoinverse $\mathbf{M}^{\dagger}$), $\mathbf{x} \mapsto \mathbf{M}^{-1}\mathbf{x}$, thereby carrying the data from intensity space to abundance space. Because $\mathbf{M}^{-1}$ is non-diagonal with substantial off-diagonal structure, this map acts geometrically as a shear combined with rotation and anisotropic rescaling of the data cloud: it mixes coordinates rather than acting on each axis independently, and it generally alters inter-marker correlations, relative population geometry, and the angles between cell clusters. Because our invariance results cover only the axis-aligned family of translations and scalings, they do not extend to this non-axis-aligned change of space. We therefore do not claim that the framework subsumes unmixing; establishing that would require demonstrating stability of the embedding under a general linear map $\mathbf{M}^{\dagger}$, a distinct and strictly stronger property than the scaling and translation invariance shown here.

This distinction also clarifies where the analysis is most naturally carried out. In abundance space each coordinate corresponds to a single marker, so optimal transport is directly interpretable: Euclidean distances between cells reflect phenotypic dissimilarity in marker expression, the transport coupling matches cells of similar phenotype, and the resulting maps and distances quantify biologically meaningful shifts in expression. In the raw intensity domain, by contrast, each axis is a linear combination of several markers and has no direct biological meaning. The method can still be applied there, but its output no longer carries this interpretation, and, without an analogous invariance (or equivariance) result under the unmixing map $\mathbf{M}^{\dagger}$, it should not be treated as equivalent to the abundance-space analysis.

More broadly, we hypothesize that many challenges conventionally addressed through preprocessing, particularly those reducible to per-marker scaling and translation, can be handled intrinsically by the proposed methodology, as we demonstrate in the Results. The framework nonetheless remains fully compatible with established pipelines and can be applied directly to data that have undergone standard quality control, unmixing, normalization, and gating, ensuring broad applicability across experimental settings.

\subsection*{Sample classification in LOT space}
In general, the task of classification is to learn from labeled training samples and assign a label to a previously unseen test sample. In our setting, we are given collections of flow cytometry distributions from multiple phenotypic classes,
\[
\{\mu^{(1)}_1, \mu^{(1)}_2, \dots\}, \quad
\{\mu^{(2)}_1, \mu^{(2)}_2, \dots\}, \quad \dots,
\]
where superscript inside (.) denotes the class label of each sample. The goal is to determine the class membership of an unknown test distribution $\mu_{\mathrm{test}}$. After applying the linearized optimal transport (LOT) embedding, each distribution is represented as a fixed-length vector in a common Euclidean space, so the problem reduces to standard vector-based classification.

Let $\{\widehat{\boldsymbol{\mu}}^{(k)}_j\}_{j=1}^{n_k}$ denote the linearized optimal transport (LOT) embeddings of training samples from class $k$, all computed with respect to a common reference distribution. We first estimate the class mean embedding
\[
\widehat{\bar{\boldsymbol{\mu}}}^{(k)} =
\frac{1}{n_k}\sum_{j=1}^{n_k}\widehat{\boldsymbol{\mu}}^{(k)}_j,
\]
and construct the centered data matrix
\[
M^{(k)} =
\bigl[
\widehat{\boldsymbol{\mu}}^{(k)}_1 - \widehat{\bar{\boldsymbol{\mu}}}^{(k)},
\;
\widehat{\boldsymbol{\mu}}^{(k)}_2 - \widehat{\bar{\boldsymbol{\mu}}}^{(k)},
\;
\dots
\bigr]
\in \mathbb{R}^{d \times n_k}.
\]
Singular value decomposition (SVD) of $M^{(k)}$,
\[
M^{(k)} = U^{(k)}\Sigma^{(k)}(V^{(k)})^\top,
\]
yields an orthonormal basis for class-specific variability. The first $r_k$ columns,
\[
U^{(k)}_{r_k} \in \mathbb{R}^{d \times r_k},
\]
define a low-dimensional subspace that captures the dominant distributional structure of class $k$.

For a previously unseen test sample with embedding $\widehat{\boldsymbol{\mu}}_{\mathrm{test}}$, we quantify its consistency with each class by computing the projection residual
\[
\mathbf{r}^{(k)} =
\Bigl(
I - U^{(k)}_{r_k}(U^{(k)}_{r_k})^\top
\Bigr)
\bigl(
\widehat{\boldsymbol{\mu}}_{\mathrm{test}} - \widehat{\bar{\boldsymbol{\mu}}}^{(k)}
\bigr),
\]
and using the squared residual norm
\[
E^{(k)} = \|\mathbf{r}^{(k)}\|_2^2
\]
as a class-specific reconstruction error. The predicted label is obtained by
\[
\hat{k} = \arg\min_k E^{(k)}.
\]

This subspace-based classifier exploits the approximately linear geometry induced by LOT embeddings and enables robust, initialization-free classification without clustering, gating, or explicit dimensionality reduction. More broadly, the same embedded representation supports alternative discriminative analyses—such as regression directions, canonical correlation analysis, or hypothesis-driven contrasts—providing a general mathematical framework for classification, visualization, and discovery that preserves a direct connection to underlying biological variation.
\subsection*{Discriminative directions and hypothesis testing in LOT space}

We consider subject-level data represented as embedded distributions,
\[
\{(\hat{\boldsymbol{\mu}}_i, y_i)\}_{i=1}^n, \qquad \hat{\boldsymbol{\mu}}_i \in \mathbb{R}^d,
\]
where each embedding $\hat{\boldsymbol{\mu}}_i$ summarizes a high-dimensional single-cell distribution, and $y_i$ denotes an associated outcome (categorical or continuous). The central objective is to identify low-dimensional directions in the embedding space that capture meaningful biological variation while suppressing nuisance effects such as technical variability or inter-subject noise.

For binary labels $y_i \in \{1,2\}$, a natural discriminative direction is obtained from the difference between class-conditional mean embeddings,
\[
\hat{\boldsymbol{\mu}}^{(k)} =
\frac{1}{|\mathcal{D}_k|}
\sum_{i \in \mathcal{D}_k} \hat{\boldsymbol{\mu}}_i,
\qquad k \in \{1,2\},
\]
and
\[
\mathbf{w}
=
\frac{\hat{\boldsymbol{\mu}}^{(2)} - \hat{\boldsymbol{\mu}}^{(1)}}
{\|\hat{\boldsymbol{\mu}}^{(2)} - \hat{\boldsymbol{\mu}}^{(1)}\|_2}.
\]
This direction defines the primary axis along which the two populations differ. Projecting each sample onto this axis,
\[
s_i
=
\mathbf{w}^\top
\bigl(\hat{\boldsymbol{\mu}}_i - \hat{\boldsymbol{\mu}}\bigr),
\qquad
\hat{\boldsymbol{\mu}} =
\tfrac{1}{2}\sum_{k=1}^2 \hat{\boldsymbol{\mu}}^{(k)},
\]
yields a one-dimensional summary that quantifies how strongly each sample aligns with the class contrast.

From an intuitive perspective, this construction measures how the \emph{same reference-centered cell displacements} shift between groups along a dominant direction. In the original distribution space, such differences may be complex and nonlinear; however, in LOT space they are captured by a single direction, converting the problem into a tractable geometric comparison. This enables direct statistical testing: for example, hypotheses of the form “the two populations differ” can be evaluated by testing for separation in the projected scores $\{s_i\}$ using standard univariate methods.

To further interpret this variation, we define a parametric path in the embedding space,
\[
\hat{\boldsymbol{\mu}}(t) = \hat{\boldsymbol{\mu}} + t\,\mathbf{w},
\]
which traces how the distribution changes along the discriminative axis. Mapping points along this path back to the original measurement space reveals how cellular populations and marker expressions deform between phenotypes, providing a direct link between statistical separation and biological interpretation.

While the mean-difference direction provides a simple and effective discriminative axis, the same LOT representation supports more general hypothesis-driven analyses. Alternative directions can be defined using regression, partial least squares, or canonical correlation to test specific biological questions. Thus, the embedding serves as a unified mathematical framework in which complex distributional differences are reduced to low-dimensional, interpretable structures that support visualization, hypothesis testing, and discovery.

\subsection*{Distribution-to-scalar regression in LOT space for MRD prediction}

As a third application, we consider the problem of predicting a clinically relevant scalar outcome from distribution-valued single-cell measurements. Given training data
\[
\{(\mu_1, q_1), (\mu_2, q_2), \dots, (\mu_n, q_n)\},
\]
where each $\mu_i$ is an empirical distribution and $q_i \in \mathbb{R}$ is an associated response, the goal is to learn a mapping that predicts the outcome for a new sample. In our setting, $q_i$ represents the percentage of blast cells among white blood cells, providing a quantitative measure of measurable residual disease (MRD). This defines a distribution-to-scalar regression problem.

To make this problem tractable, each distribution $\mu_i$ is first embedded into a finite-dimensional vector space using the LOT framework, yielding $\widehat{\boldsymbol{\mu}}_i \in \mathbb{R}^d$. This step converts complex population-level variation into structured coordinates that can be directly used in regression. We then model the relationship between embedding and outcome using partial least squares (PLS), which identifies directions in the embedding space that are maximally associated with the response. Specifically, each latent component takes the form
\[
t_i = \mathbf{w}^\top \widehat{\boldsymbol{\mu}}_i,
\]
where the direction $\mathbf{w}$ is chosen to maximize covariance with the MRD values,
\[
\max_{\|\mathbf{w}\|_2=1} \left( \frac{1}{n} \sum_{i=1}^n (\mathbf{w}^\top \widehat{\boldsymbol{\mu}}_i)\, q_i \right)^2.
\]
Iteratively extracting such components yields a low-dimensional representation that captures the most predictive modes of variation. The final prediction model is given by
\[
\hat{q}_i = \beta_0 + \sum_{k=1}^K \beta_k \, t_i^{(k)},
\]
with $K$ chosen by cross-validation.

From an intuitive perspective, this approach identifies directions in LOT space along which shifts in the embedded cell populations are most strongly associated with disease burden. Because the embedding tracks how the same reference cells are displaced across samples, these predictive directions correspond to structured changes in population composition and marker expression. Thus, the regression model does not simply fit high-dimensional features, but rather isolates biologically meaningful modes of distributional deformation that relate directly to MRD. More broadly, this formulation enables hypothesis-driven analysis, where one can test whether specific directions of population change are predictive of clinical outcomes, providing an interpretable link between single-cell dynamics and disease progression.

\end{document}